\documentclass{llncs}
\usepackage{makeidx,graphicx}  

\begin{document}

\mainmatter              

\title{Systems Dynamics or Agent-Based Modelling for Immune Simulation?}

\author{Grazziela P. Figueredo \and Uwe Aickelin \and Peer-Olaf Siebers}

\authorrunning{G. Figueredo et al.}   
%
\institute{Intelligent Modelling and Analysis Research Group, School of Computer Science, \\ The University of Nottingham, NG8 1BB, UK \\ gzf, uxa, pos@cs.nott.ac.uk}



\maketitle

\begin{abstract}
\noindent

In immune system simulation there are two competing simulation approaches: System Dynamics Simulation (SDS) and Agent-Based Simulation (ABS). In the literature there is little guidance on how to choose the best approach for a specific immune problem. Our overall research aim is to develop a framework that helps researchers with this choice. In this paper we investigate if it is possible to easily convert simulation models between approaches. With no explicit guidelines available from the literature we develop and test our own set of guidelines for converting SDS models into ABS models in a non-spacial scenario. We also define guidelines to convert ABS into SDS considering a non-spatial and a spatial scenario. After running some experiments with the developed models we found that in all cases there are significant differences between the results produced by the different simulation methods.
\end{abstract}

\section{Introduction}
\label{Introduction}

Simulation presents paradigms that allow us to build models for various problem domains. Some of the important simulation approaches are System Dynamics Simulation (SDS) and Agent-Based Simulation (ABS). SDS is a continuous simulation approach that uses stocks, flows and feedback loops as concepts to study the behaviour of complex systems~\cite{Forrester:1969}. The models in SDS consist of a set of differential equations that are solved for a certain time interval~\cite{Macal:2010}. ABS, on the other hand, is a modelling technique that employs autonomous agents that interact with each other. The agents' behaviour is described by rules that determines how they learn, interact with each other and adapt. The overall system behaviour is given by the agents individual dynamics as well as their interactions. Table~\ref{Tab:DifferencesSDSABS} shows a summary of ABS and SDS features considering their main aspects and differences.

\begin{table}[htpb]
\caption{Main differences between SDS and ABS (obtained
from~\cite{Schieritz:2003}).}

\begin{center}
\begin{tabular}{l l l l l}
\hline
Feature           &  &SDS            & &  ABS            \\
\hline

Perspective       &  &top-down &       &  bottom-up      \\

Building block    &  &feedback loop &  &  agent          \\

Unit of analysis  & &system structure && agent's rules   \\


Level of modelling && aggregate       && individual      \\


System structure   && fixed           && not fixed       \\


Time handling      && continuous      && discrete \\
\hline

\end{tabular}

\label{Tab:DifferencesSDSABS}
\end{center}
\end{table}

SDS is widely applicable at a high level of abstraction. ABS, on the other hand, is a paradigm that can be used at any level of abstraction, including those levels covered by SDS.  As there is an intersection, a range of simulation problems can be solved either by SDS or ABS. In~\cite{Thorne:2007}, the authors state that ABS is ideal for tissue patterning events because it explicitly represents individual cells in space and time. Moreover, ABS indicates how the tissue behaviour emerges from the interactions of individual cells. On the other hand, ABS requires computational power and may produce large sets of data, which could be difficult to analyse~\cite{Thorne:2007}. In addition, ABS requires all system's properties to be modelled discretely. SDS, however, deals with continuum approximations. For the simulation of biological systems, therefore, both approaches are useful and should be selected carefully according to the research question to be addressed.

We believe that these two approaches can be very useful for the simulation of parts of the immune system, based on the findings of~\cite{Thorne:2007}. However, there is still little knowledge on how to determine the best approach for a given immune problem. Moreover, little is known concerning the comparison of SDS and ABS  for simulation in immunology. Hence, our study aims to establish a framework to help with the choice between SDS and ABS approaches for immune system problems.

In previous work~\cite{Figueredo:2010} we compared the use of ABS and SDS for modelling non-spatial static agents' behaviour in an immune system ageing problem. By static we mean that there is no movement or interactions between the agents. We concluded that for these types of agents, it is preferable to use SDS instead of ABS. When contrasting the results of both simulation approaches, we saw that SDS is less complex and takes up less computational resources, producing the same results as those obtained by the ABS model. In addition, SDS is more robust when the number of cells increase considerably. There were cases where there was not enough computational resources to run the ABS.

More recently, we used case studies which included interactions between tumour cells and immune effector cells, reviewed in~\cite{Eftimie2010}. We began with the simplest (single equation) models for tumor growth and proceed to consider two-equation models involving effector and tumour cells. We used mathematical models as basis for both ABS and SDS. The idea was to check if the results are similar and if we can use SDS and ABS for our case studies interchangeably. In our experiments we obtained different outputs from the ABS and the SDS. This is due to the fact that SDS is a deterministic method while ABS is a stochastic method. To proceed with our tests, we considered tumour cells growth together with their interactions with general immune effector cells. In this case, there were also differences in the output because the effector cell numbers change continuously in the SDS, while for the ABS, they change in a discrete pattern. For example, in the SDS it is possible to consider cases where there are 0.5 cells, while for the ABS it is either 0 or 1 cell (agent).

To advance our study, we have two research objectives. Nuno {\it et al.} \cite{Nuno:2007} mention that most of what has been done in simulation of the immune system is based on differential equations, which can be easily implemented using SDS, as long as feedback loops are expressed in the equations. We believe that the conversion of current well established mathematical or SDS models into ABS would be a first step to investigate individual behaviour and emergence on the existing models. Hence, the first objective is to develop and test our own set of guidelines for converting SDS models into ABS models. For this we will use a non-spatial model involving interactions between the immune system and cancer. Our second objective is to define guidelines to convert ABS into SDS considering a non-spatial and a spatial scenario.

The remainder of the paper is organized as follows. Section~\ref{RelatedWork} presents a literature review of works comparing SDS and ABS for biological problems. In Section~\ref{SDSToABS}, we address our first research objective by presenting an example of conversion from SDS to ABS as well as the results comparison. In Section~\ref{ABSToSDS}, we present two simulation models implemented in ABS and their conversion into SDS models and compare their results. Finally, in Section~\ref{Conclusions}, we draw the conclusions and present ideas to continue our study.

\section{Related Work}
\label{RelatedWork}

In this section, we describe the literature concerned with the comparison between ABS and SDS for biological problems. We found that there is hardly any literature comparing the two approaches for immune simulation.

Wayne {\it et al. }\cite{Wayne:2004} show the application of both SDS and ABS to simulate non-equilibrium cellular ligand-receptor dynamics over a broad range of concentrations. They concluded that both approaches are powerful tools and are also complementary. In their case study, they did not indicate a preferred paradigm, although they state that intuitively SDS is an obvious choice when studying systems at a high level of aggregation and abstraction. On the other hand, SDS is not capable of simulating receptors and molecules and their individual interactions, which can be done with ABS.

Rahmandad and Sterman ~\cite{Sterman:2008} compare the dynamics of a stochastic ABS model with those of the analogous deterministic compartment differential equation model for contagious disease spread. The authors convert the ABS into a differential equation model and examine the impact of individual heterogeneity and different network topologies. The deterministic model yields a single trajectory for each parameter set, while stochastic models yield a distribution of outcomes. Moreover, the differential equation model and ABS dynamics differ for several metrics relevant to public health. The response of the models to policies can also differ when the base case behaviour is similar. Under some conditions, however, the differences in means are small, compared to variability caused by stochastic events, parameter uncertainty and model boundary.

As we mentioned before, in our previous work~\cite{Figueredo:2010}, we compared SDS and ABS for a naive T cell output model. We had a scenario where the agents had no interactions and SDS and ABS produced similar outputs, although SDS takes up less computational resources. We concluded, therefore, that SDS is more suitable.  More recently, we used case studies which included interactions between tumour cells and immune effector cells. We wanted to know if the results would be similar and if we can use SDS and ABS for our case studies interchangeably. In our experiments, the stochastic behaviour of the agents made the output from ABS different from the SDS output. Moreover, there were differences in the outcomes due to the continuous character of SDS contrasted to the discrete behaviour of ABS.

Macal ~\cite{Macal:2010} shows how to translate a SDS into an equivalent time-stepped, stochastic agent-based simulation. Probabilistic elements in the SDS model were identified and translated into probabilities that were used explicitly in the ABS model. The author uses as an example the SIR model proposed by Kermack and McKendrick~\cite{Kermack:1927}. This model was built to understand and predict the spread of epidemics. In the model, the population is divided in three groups of individuals, susceptible (S), infected (I) and recovered (R). To convert the model from SDS to ABS, the author considers two agent-based formulations. Model 1 was defined as a {\it ``naive ABS model, because it provides no additional information or implementation advantages over the SDS model''}. There is a set of agents containing a state (S, I or R), which is the only information dynamically updated. The author claims that Model 1 produces exactly the same results for the numbers of S, I and R over time as does the SDS model for a fixed-time step, $\Delta t$ of length one. Model 2, on the other hand, is fully individual-based agent model and provides additional information over the SDS model. As for example, in some of the ABS simulation runs of Model 2 an epidemic does not occur. In a significant number of cases, the number of contacts and the number of infected individuals (I) is not large enough to spread the infection. These cases occur because of the agent's probabilistic rules. Hence, Model 2 presents similar results from SDS, but not exactly the same because of the runs where there was not epidemic. The author, therefore, concludes that the ABS model is able to provide additional information over what the SDS model provides given its stochastic nature.

In this paper we want to use examples from the immune area to convert from SDS to ABS, similarly to how it was done in~\cite{Macal:2010}. Moreover, we define explicit guidelines for this conversion and bring some other questions concerned with the choice of SDS and ABS. In the next section, we perform the SDS to ABS conversion for a mathematical model of the interactions between tumour cells and effector cells.

\section{From SDS to ABS}
\label{SDSToABS}

In this section we address our first research objective by using a mathematical model to build a non-spatial SDS model and then convert it into an ABS model.

\subsection{The Mathematical Model}

The mathematical model we use to build our SDS was obtained from~\cite{Kirschner:1998}. The model's equations illustrate the non-spatial dynamics between effector cells (E), tumour cells (T) and the cytokine IL-2 ($I_L$). The model is described by the following differential equations:

\begin{equation}
\label{Eq:1}
 \frac{dE}{dt} = cT - \mu_2E + \frac{p_1EI_L}{g_1+I_L} + s1,
\end{equation}

Equation~\ref{Eq:1} describes the rate of change for the effector cell population E~\cite{Kirschner:1998}. Effector cells grow based on recruitment ($cT$) and proliferation ($\frac{p_1EI_L}{g_1+I_L}$). The parameter $c$ represents the antigenicity of the tumour cells (T)~\cite{Kirschner:1998,Arciero:2004}. $\mu_2$ is the death rate of the
effector cells.  $p_1$ and $g_1$ are parameters used to calibrate the recruitment of effector cells and $s1$ is the treatment that will boost the number of effector cells.

\begin{equation}
\label{Eq:2}
 \frac{dT}{dt} = a(1 - bT) - \frac{a_aET}{g_2 + T},
\end{equation}

Equation~\ref{Eq:2} describes the changes that occur in the tumour cell population T over time. The term $a(1 - bT)$ represents the logistic growth of T ($a$ and $b$ are parameters that define how the tumour cells will grow) and $\frac{a_aET}{g_2 + T}$ is the number of tumour cells killed by effector cells. $a_a$ and $g_2$ are parameters to adjust the model.

\begin{equation}
\label{Eq:3}
 \frac{dI_L}{dt} = \frac{p_2ET}{g_3 + T} - \mu_3I_L + s2.
\end{equation}

The IL-2 population dynamics is described by Equation~\ref{Eq:3}. $\frac{p_2ET}{g_3 + T}$ determines IL-2 production using parameters $p_2$ and $g_3$. $\mu_3$ is the IL-2 loss. $s2$ also represents treatment. The treatment is the injection of IL-2 in the system.

\subsection{The SDS Model}

The SDS model contains three stock variables, tumour cells, effector cells and IL-2. The stock of effector cells, described by Equation~\ref{Eq:1}, is changed by the recruitment of new effector cells, according to the number of tumour cells, death, proliferation and treatment (insertion of new effector cells). The conversion of Equation~\ref{Eq:1} into a stock and flow diagram can be seen in Figure 1(a). The number of
tumour cells is changed by its natural proliferation and death as well as by the number of cells killed by effector cells (Figure 1(b)). IL-2 stock changes with the production of new IL-2 molecules from effector cells (the production also depends on the number of tumour cells), loss and treatment (insertion of IL-2) (Figure 1(c)). The final SDS stock and flow diagram is depicted in Figure 2. We obtain this diagram by associating the flows with the stocks that will influence them. This information is obtained by to the equations of the mathematical model. For example, we know that the number of tumour cells killed is dependent on the number of effector cells. Hence, we have to add the number of effector cells on the mathematical equation of the flow $TumourKilledByEffectorCells$.

\begin{figure}[!htpb]
  \resizebox{12cm}{!}{\includegraphics{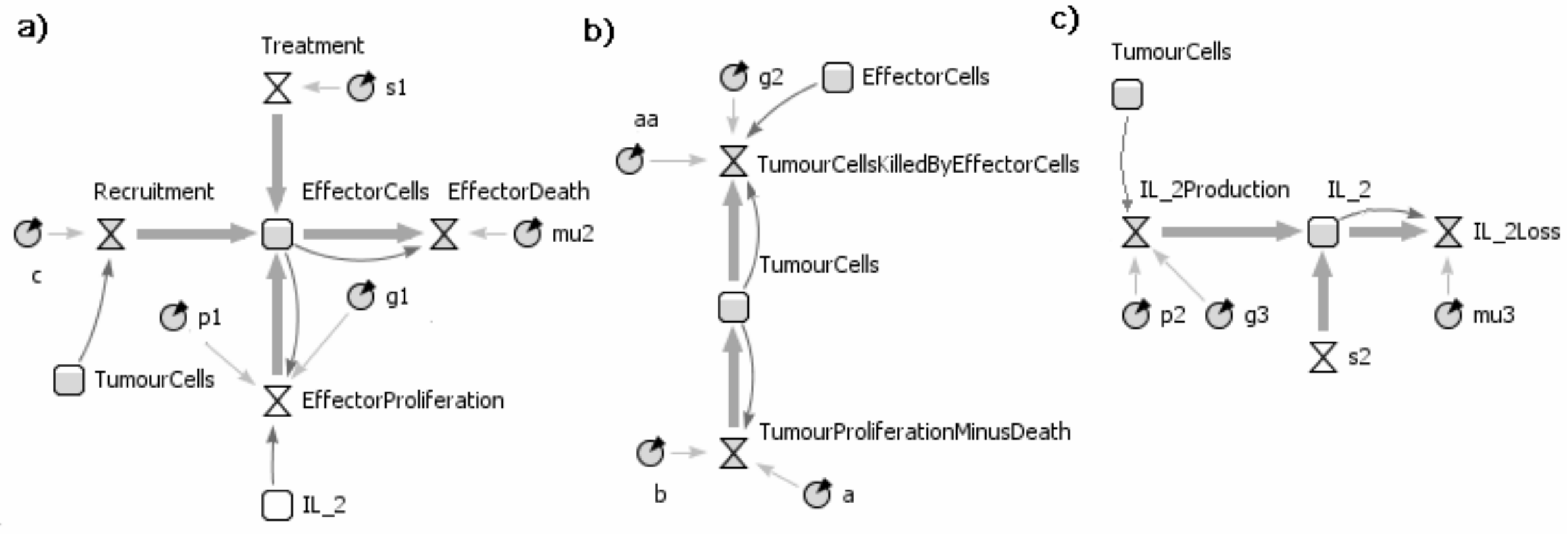}}
 \label{fig:EquationsToSDS}
 \caption{Equations 1, 2 and 3 converted into stock and flow diagrams (squares = stocks, hourglasses = flows, circles = parameters and arrows = information flows).}
\end{figure}

\begin{figure}[!htpb]
 \begin{center}
  \resizebox{11cm}{!}{\includegraphics{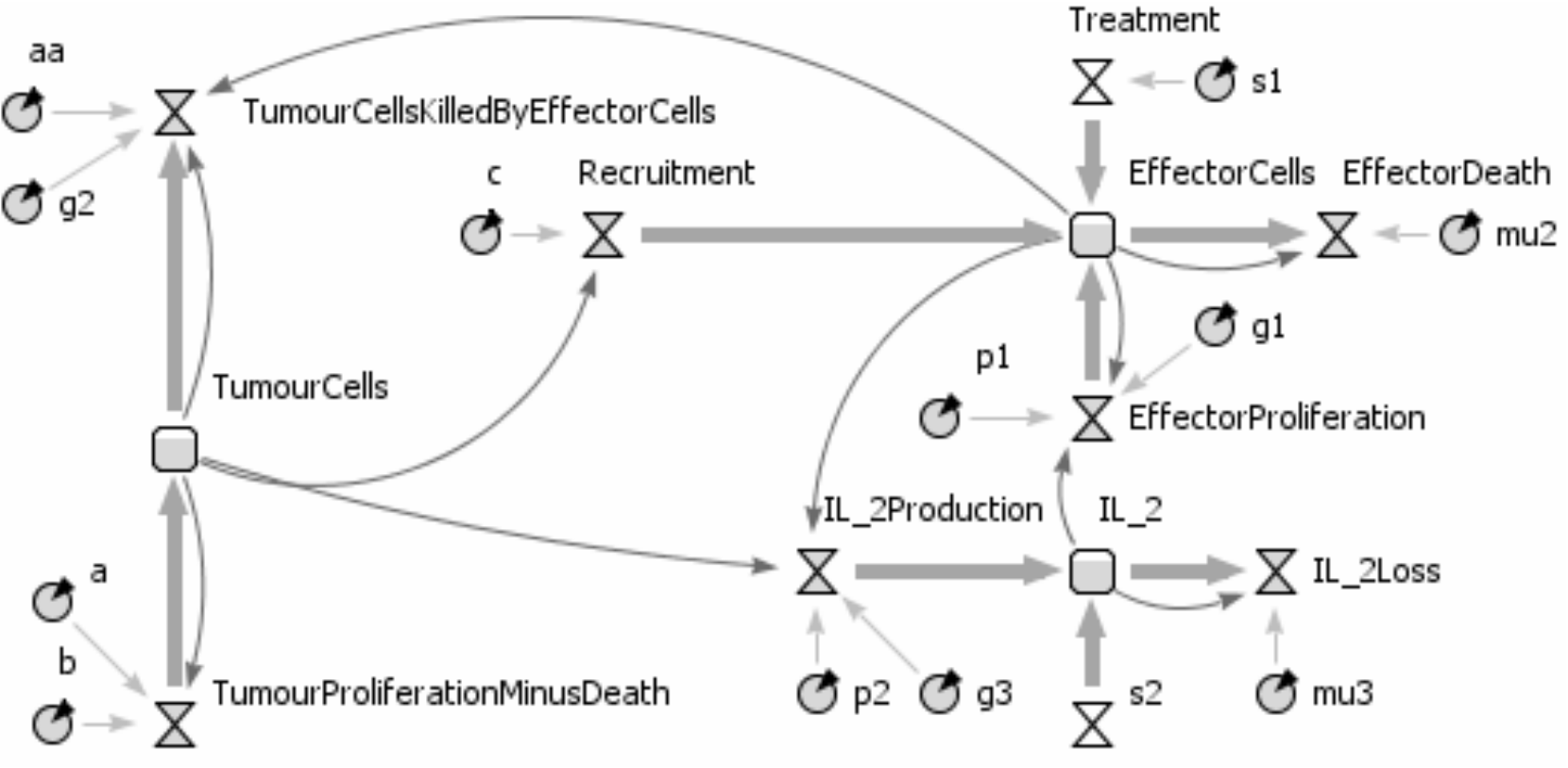}}
 \end{center}
 \label{fig:SDSmodelThreeEquation}
 \caption{SDS diagram for the three-equation mathematical model.}
\end{figure}

\subsection{The ABS Model}

In order to convert the SDS into an ABS model, we propose the following steps:

{\bf 1. Identify the possible agents}. To do so we use some characteristics defined in~\cite{Macal:2005}. An agent is: (1) self-contained, modular, and uniquely identifiable individual; (2) autonomous and self-directed; (3) a construct with states that varies over time and (4) social, having dynamic interactions with other agents that impact its behaviour. In the SIR model discussed in~\cite{Macal:2010}, the author defines two possible ABS. However, in our case we believe that the best way to address the modelling problem is by defining one ABS with three groups of agents, similar to the implementation of Model 2 in~\cite{Macal:2010}. Our agents will be corresponding to the stocks in the SDS model. Hence, the populations of agents are the effector cells, tumour cells and IL-2s. It is important to mention that we converted stocks into agents for this specific problem. However, there are cases where stocks might not be agents. For instance, in our simulation experiments performed for the naive T cell output model in~\cite{Figueredo:2010}, the stocks were states of only one agent representing a T cell.

{\bf 2. Identify the behaviour of each agent}. As we are building the ABS model from an SDS model, the agent's behaviours will be determined by mathematical equations converted into rules. The behaviour of each agent can be seen in Table~\ref{Tab:ABSModelThreeEquation}. Each agent has two different types of behaviours: reactive and proactive behaviours. The reactive behavior occurs when the agents perceive the context in which they operate and react to it appropriately. The proactive behaviour describes the situations when the agent has the initiative to identify and solve an issue in the system.

\begin{table}[h]
\caption{Agents' conceptual model (informed by Equations~\ref{Eq:1}, \ref{Eq:2} and \ref{Eq:3}).}

\begin{center}

\begin{tabular}{l|l|l|l}
\hline
Agent          & Parameters             & Reactive behaviour            &  Proactive behaviour            \\
\hline \hline

Effector Cell  & (1)Death Rate          & (1)Dies with age              & Reproduces                      \\
               & (2)Reproduction Rate   & (2)Is recruited               &                                 \\
               &                        & (3)Is injected as treatment   &                                 \\

\hline

Tumour Cell    & (1)Death Rate          &  (1)Dies killed by effector cells & Reproduces         \\
               & 2)Reproduction Rate    &  (2)Dies with age                                      \\
\hline

IL-2           &                         &  (1)Is produced   &\\
               &                         &  (2)Is lost       &\\
               &                         &  (3)Is injected   &\\

\hline
\end{tabular}

\label{Tab:ABSModelThreeEquation}

\end{center}

\end{table}

{\bf 3. Build the agents}. Based on the conceptual model derived from step 2 (Table 2) we have developed some state charts, one for each agent type (Figure 3). In the state charts we model states and state transitions.

\begin{figure}[!htpb]
 \begin{center}
  \resizebox{9cm}{!}{\includegraphics{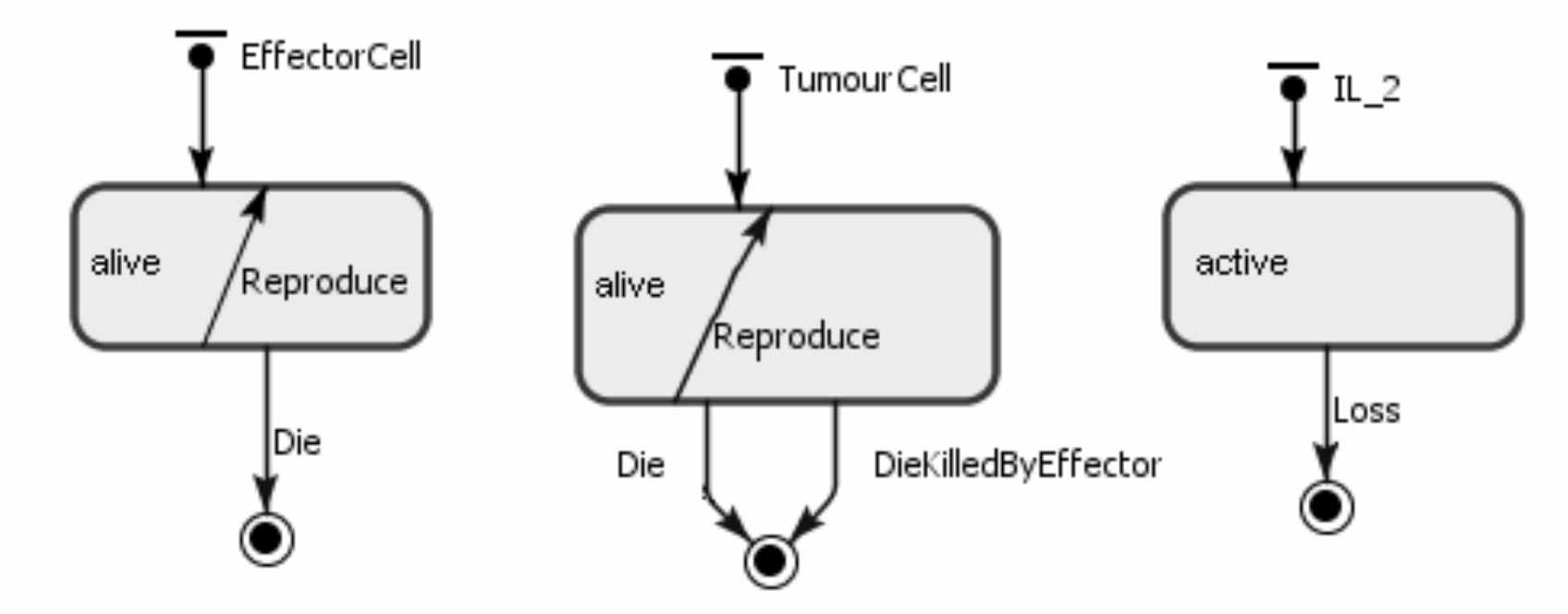}}
 \end{center}
 \label{fig:ABSModelThreeEquation}
 \caption{ABS state charts for the agents of the mathematical model (squares = agent's states, arrows = transitions and filled circles with a ring = final states.}
\end{figure}

\subsection{Results Comparison}

First we validated our SDS model by comparing its outputs to the outputs produced by the mathematical model derived from~\cite{Kirschner:1998}. Both produced very similar results. Here we validate our ABS model by comparing its outputs to the outputs produced by the SDS model (i.e. our base model for the comparison). We ran the simulations on an $Intel®$ $Core^{TM}$ Duo CPU 2GHz and 2GB RAM. We simulated a period of 400 days. As ABS is a stochastic simulation method we had to conduct several replications. We decided to run 50 replications and calculated the mean values for the outputs. The results for both simulations are shown in Figure 4. From the graphs it is very obvious that the results are very different for the two different simulation approaches.

In the SDS results, tumour cells decrease as effector cells increase, following a predator-prey trend curve. As SDS works with continuous numbers, by 200 days tumour cells and effector cells asymptotically tend to zero. This allows for the populations to increase again, as they never reach zero. On the other hand, for the ABS, the number of effector cells decreases until zero and therefore, proliferation stops. Further investigation needs to be done to see if this is a unique case of if we produce similar results when converting related problems. Moreover, we intend to modify the SDS so that it will consider only discrete numbers of effector cells and investigate the results.

\begin{figure}[!htpb]
 \begin{center}
  \resizebox{13cm}{!}{\includegraphics{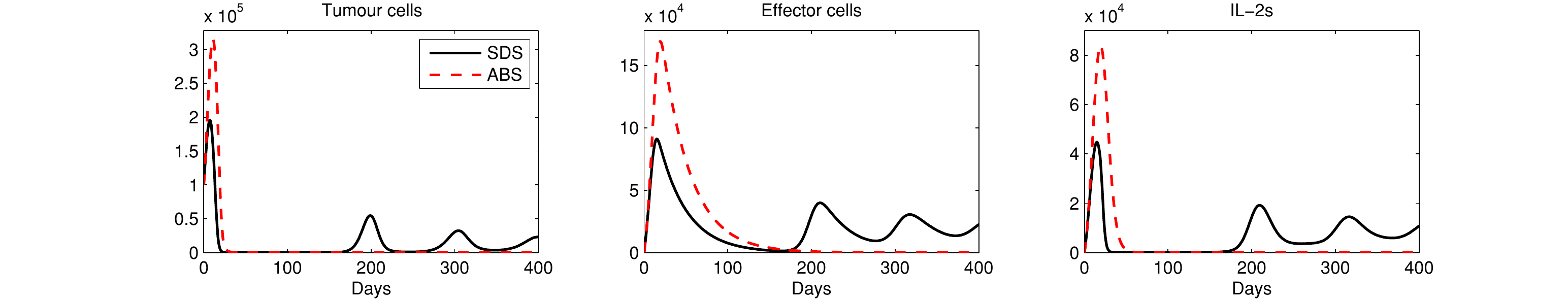}}
 \end{center}
 \label{fig:ResultsThreeEquation}
 \caption{SDS and ABS results.}
\end{figure}

\section{From ABS to SDS}
\label{ABSToSDS}

In this Section we address our second objective, which is to define some guidelines to convert a simple ABS model into an SDS model. Next, we present the simulation scenario description.

\subsection{Simulation Problem Description and Conceptual Modelling}

Effector cells are recruited after a tumour is detected in the organism. Their role is to search and kill tumour cells inside the tumour. Tumour cells reproduce and die with age or are killed by effector cells. For the simulation, we defined the following agents (classes): tumour cell and effector cell. The conceptual model of our agents is given in Table 3. For each agent we present the class specification (parameters and behaviours).

\begin{table}[h]
\caption{Agents' conceptual model.}
\begin{center}
\begin{tabular}{l|l|l|l}
\hline
Agent          & Parameters              &  Reactive behaviour       & Proactive behaviour                \\
\hline \hline

Effector Cell  & (1)Death Rate           &  (1)Dies with age         & (1)Reproduces             \\

               & (2)Reproduction rate    & (2)Is recruited           & (2)Kills tumour cells       \\

\hline

Tumour Cell    & (1)Death Rate           &(1)Dies with age            &  (1)Reproduces              \\

               & (2)Reproduction rate    &  (2)Is killed by effector cells &                 \\
\hline
\end{tabular}
\label{Tab:ABSAgentsModel1}
\end{center}
\end{table}

For our experiments, we considered two ABS implementations. The first implementation does not consider cellular movement. The second ABS model allows for effector cell movement. We decided to add space as an additional variable for our simulations because it makes a better match between the simulation and the real world.

\subsection{ABS  for Model 1}

From the conceptual model we have built the ABS model shown in Figure 5. The figure shows the state charts for effector cells and tumour cells agents.

\begin{figure}[htpb]
 \begin{center}
  \resizebox{9cm}{!}{\includegraphics{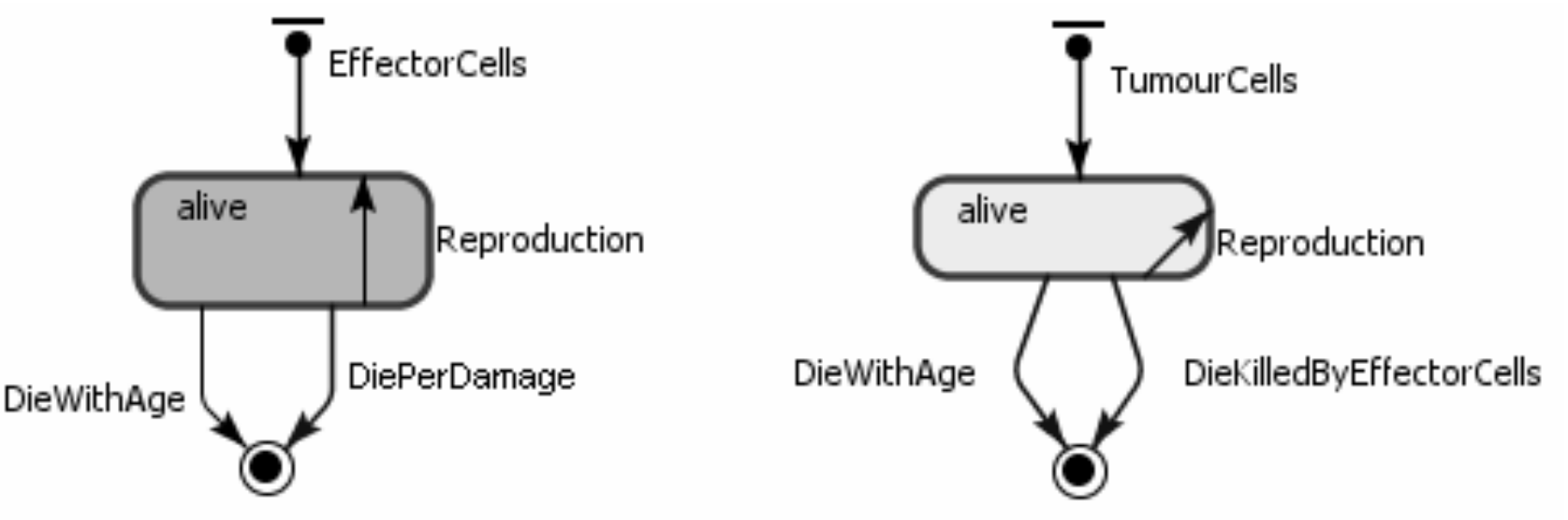}}
 \end{center}
 \label{fig:TwoEquationABS}
 \caption{ABS diagram for Model 1.}
\end{figure}

\subsection{SDS for Model 1}

For converting the ABS model into an SDS model, we propose the following steps:

{\bf 1. Identify the system structure}. First we have to recognize the system structure and assume a high level of aggregation for the objects being modelled. It is necessary to generalise from the specific events and consider patterns of behaviour that characterise the situation. The cells, therefore, will no longer respond individually. The simulation outcome will be given by the collection of cells and its dynamics as a group. In our case, we have two cell populations (aggregations). Looking at the ABS diagram of Figure 5, the tumour cell population changes with time by reproduction, natural death and death caused by immune cells. The second population are the effector cells. They die with age or apoptosis/damage and reproduce. We know that the effector cell population negatively impacts the amount of tumour cells because effector cells kill tumour cells with time. The reproduction of effector cells increases as the number of tumour cells increase. In addition, as effector cells kill tumour cells they get damaged. Therefore, the tumour cell population impacts the effector cells population in both positive and negative ways.

{\bf 2. Identify the stocks in the system}. Stocks are physical entities which can accumulate over time. In our example, we defined as stocks the effector cells and tumour cells.

{\bf 3. Define the stocks and their flows}. Having the stocks (step 2) and the information about the structure of the model (step 1) we can depict how each stock is changed over time by the flows and the information about how a stock would influence a flow. The effector cells stock will be decreased by death and apoptosis. Moreover, this will be increased by proliferation. The number of effector cells in the system also influences the number of cell's death, proliferation and apoptosis. Therefore, we will have a stock and flow diagram shown in Figure 6(a). The same happens with the tumour cell stock, which is changed by proliferation and death (Figure 6(b)).

\begin{figure}[hb]
 \begin{center}
  \resizebox{11cm}{!}{\includegraphics{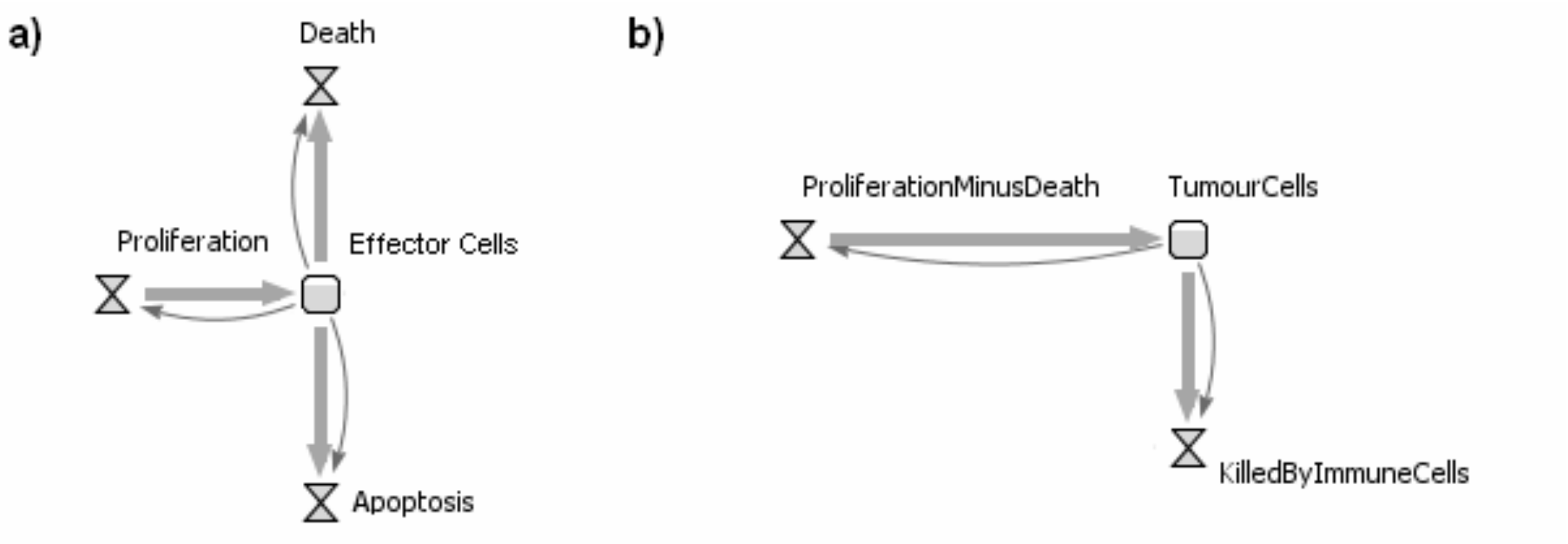}}
 \end{center}
 \label{fig:Model1StocksFlows}
 \caption{Stock and flow diagram for effector cells (a) and tumour cells (b).}
\end{figure}

{\bf 4. Define the final stock and flow diagram}. After defining the diagrams for each stock, it is necessary to go back to the system structure and define how the stocks will interact or influence each other. As we mentioned before, tumour cells impact on the proliferation and death of effector cells, and effector cells influence the growth of tumour cells (Figure 7).

\begin{figure}[htpb]
 \begin{center}
  \resizebox{9.5cm}{!}{\includegraphics{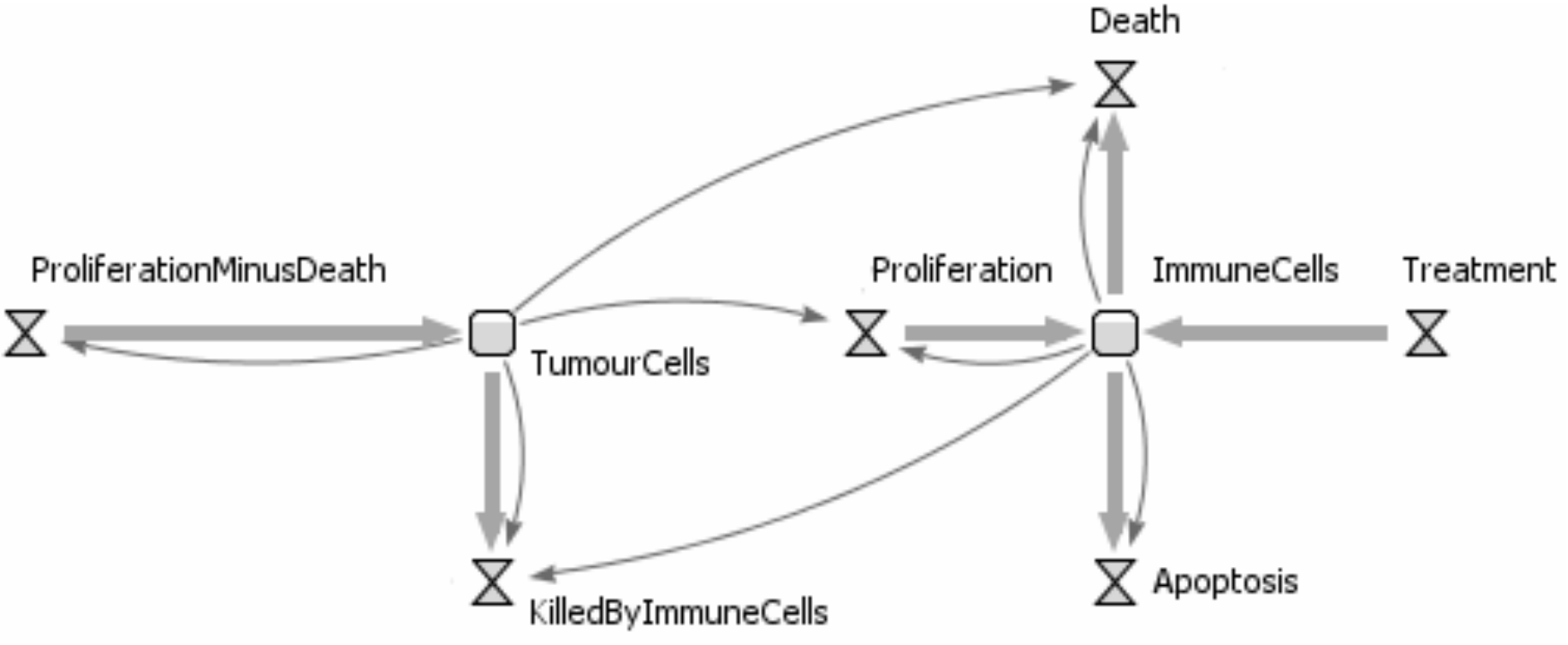}}
 \end{center}
 \label{fig:twoEquationSDS}
 \caption{SDS diagram for Model 1.}
\end{figure}

{\bf 5. Define the mathematical model}. For SDS, a set of mathematical equations is necessary to describe how the stocks will change over time. By looking at the diagram of Figure 7, the interactions between tumour cells and immune effector cells can be defined by the equations:

\begin{equation}
\frac{dT}{dt} = p_T(T)- d1_T(T) - d2_T(T,E)
\label{Eq:Tumour}
\end{equation}

\begin{equation}
\frac{dE}{dt} = p_E(T,E) - d_E(T,E) - a_E(E),
\label{Eq:Effector}
\end{equation}

where: $T$ is the number of tumour cells, $E$ is the number of effector cells, $p_T(T)- d1_T(T)$ is the growth of tumour cells ($proliferation - natural\_death$), $d2_T(T,E)$ is the number of tumour cells killed by effector cells, $p_E(T,E)$ is the proliferation of effector cells and $a_E(E)$ is the death (apoptosis) of effector cells.

For SDS the information provided by the ABS is not enough, because we do not have the equations and rates  defining the growth or death of each cell population. Therefore, to continue building the model we need extra information. For example, a data set or a well established model that describes mathematically how the system changes over time would be necessary. For our case study, we used the data and equations defined in~\cite{Kuznetsov:1994}:

\begin{equation}
p_T(T)- d1_T(T) = Ta(1-bT),
\end{equation}

\begin{equation}
d2_T(T,E)=TE,
\end{equation}

\begin{equation}
p_E(E,T)= \frac{pTE}{g+T},
\end{equation}

\begin{equation}
d_E(E,T)=mTE,
\end{equation}

\begin{equation}
a_E(E)= dE,
\end{equation}

Where: $a = 1.636$, $b = 0.002$, $d = 0.3743$, $g = 20.19$, $m = 0.00311$ and $p = 1.131$. We got these values from \cite{Eftimie2010} and used these parameters on the ABS. The models validation was based on the results shown in~\cite{Eftimie2010}.

\subsection{Results for Model 1}

We validated our simulation models by comparing its outputs to the outputs produced by the mathematical model derived from~\cite{Kuznetsov:1994}, as we have done in Section 3.4. We simulated Model 1 for a period of 100 days using both approaches. We ran 50 replications for the ABS and calculated the mean values for the outputs. The results are shown in Figure 8. In the figure, we plotted the results for the first 60 days, where the simulations reach a steady-state.The outputs produced by the SDS model are the same as the mathematical model and different from the results produced by the ABS model. The growth of tumour cells happens faster in the ABS. Another difference is in the decay of effector cells. By 20 days, there is an increase in the number of effector cells for the SDS, while in the ABS it does not increase. We believe this is due to the continuous character of SDS compared to ABS. Therefore, once effector cells population decreases to zero in the ABS, it does not increase again. The SDS the effector cells decrease to zero asymptotically. As they do not reach the value zero, they grow back again, according to the mathematical definitions (Equations~\ref{Eq:Tumour} and~\ref{Eq:Effector}).

\begin{figure}[htpb]
\begin{flushleft}
 \resizebox{10cm}{!}{\includegraphics{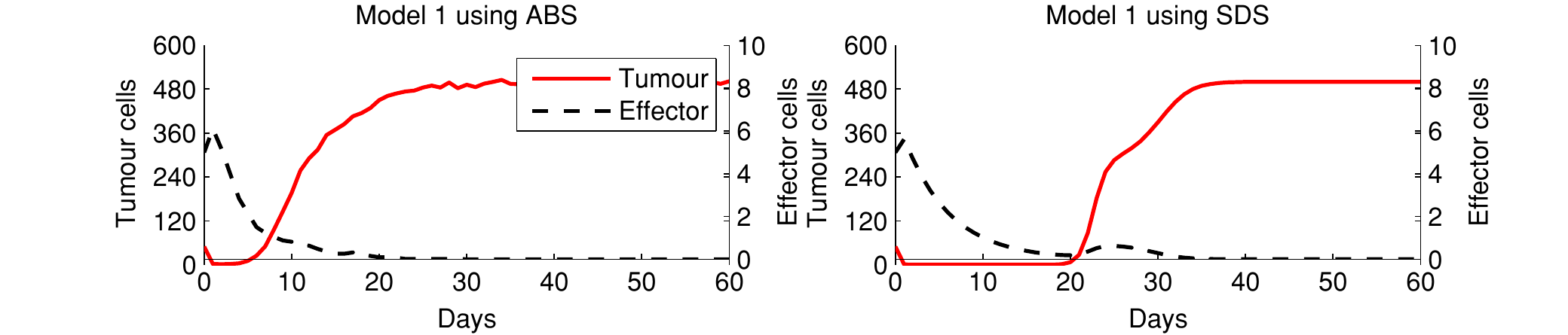}}
 \label{fig:ResultsModel1}
 \caption{ABS and SDS results for Model 1.}
 \end{flushleft}
\end{figure}

\subsection{ABS for Model 2}

As Model 1 is non-spatial, the effector cells do not move to reach a tumour cell. It impacts on the results of the simulations because tumour cells die in a rate that considers the entire population of effector cells. If we want to simulate a scenario closer to reality, a certain effector cell $E_{c_i}$ has to move towards a tumour cell $T_{c_i}$ and kill it. The remaining effector cells in the population will not have any impact on the death of $T_{c_i}$. Therefore, we decided to introduce space in Model 2. Our goal is to verify the differences in the model's behaviour over time, compared to Model 1. We also want to investigate how the movement of effector cells would impact on the SDS model development. The agents of Model 2 will be the same considered in Model 1 (Figure 5), as well as their behaviours (Table 3). However, we added an additional rule to the effector cell agents: they have to move towards the closest tumour cell and then kill it.

\subsection{SDS for Model 2}

Although the stock and flow diagram for Model 2 remained the same as for Model 1 (Figure 7), we had to adapt the mathematical model to include effector cell movement. At each time step in the simulation, the maximum number of tumour cells killed will be equal to the number of effector cells. Hence, Equation~\ref{Eq:Tumour} will be replaced by Equation~\ref{Eq:TumourMovement}:

\begin{equation}
\frac{dT}{dt} = p_T(T)- d1_T(T) - d2_T(E)
\label{Eq:TumourMovement}
\end{equation}

where: $d2_T(E)=E$

\subsection{Results for Model 2}

The simulation results for Model 2 are shown in Figure 9. For the ABS results, we display the mean value of 50 runs. The outcomes for both simulations strategies is similar, although the mean number of effector cells in the ABS is 40\% higher. With these results we show that for this example, the cellular movement can be also simulated in the SDS.

\begin{figure}[htpb]

  \resizebox{11cm}{!}{\includegraphics{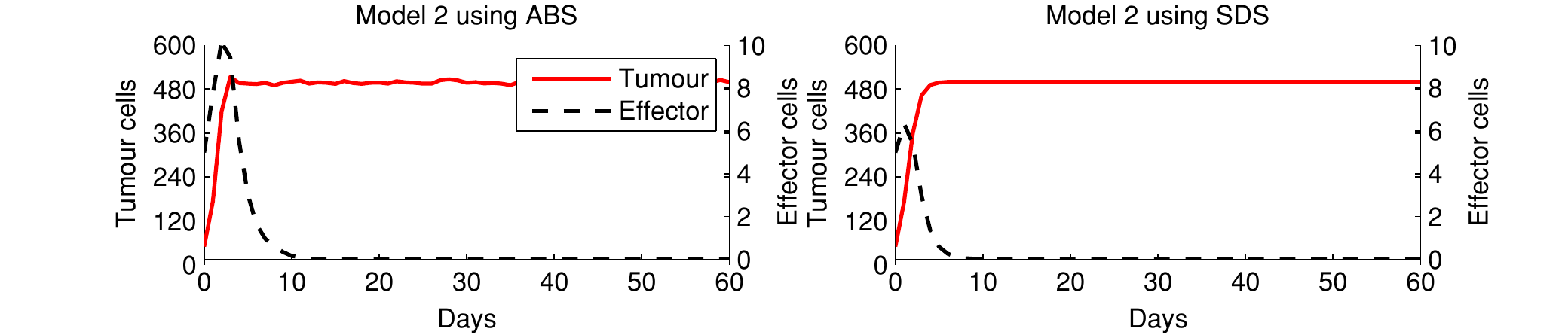}}

 \label{fig:ResultsModel2}
 \caption{ABS and SDS results for Model 2.}
\end{figure}

\section{Conclusions}
\label{Conclusions}

In the literature there is little guidance on how to choose between SDS and ABS for a specific immune problem. Our overall research aim is to develop a framework that helps with this choice. In this work we investigated the question if it is possible to easily convert simulation models between these two approaches. As there are no explicit guidelines available from the literature, we developed and tested our own set of guidelines for converting SDS models into ABS models and vice-versa. While in the first case we only considered a non-spatial scenario, in the latter case we looked at a non-spatial and a spatial scenarios.  Our results showed that it is possible to obtain an ABS model based on the information inherited in a SDS model. However, the outcomes for the two approaches were different. For the conversion of ABS into an SDS we realized that, for our case study, extra information was necessary to build some of the mathematical equations required for the SDS model. Hence, we had to use a mathematical model established in the literature to continue building the simulation. We obtained different outcomes in our non-spatial scenario. When we added spatial dimension to the agents in the ABS we still managed to adapt the SDS model, with similar results. As future work, we intend to investigate further the reasons why our conversion from SDS into an ABS produced different results.
We will implement further models to see if there is a systematic error behind these mismatches.

\bibliographystyle{splncs}
\bibliography{tese}

\end{document}